\documentclass{iau}

\usepackage{amsmath}
\usepackage{graphicx}
\usepackage{multirow}

\begin{document}

\lefttitle{Sung Kei Li}
\righttitle{Massive stars as gravitationally lensed transients – Insights on the high-mass initial mass function}

\jnlPage{1}{7}
\jnlDoiYr{2021}
\doival{10.1017/S1743921326106152}

\aopheadtitle{Proceedings IAU Symposium}
\editors{A. Wofford,  N. St-Louis, M. Garcia \&  S. Simón-Díaz, eds.}

\title{Massive stars as gravitationally lensed transients – Insights on the high-mass initial mass function}

\author{Sung Kei Li}
\affiliation{Department of Physics, The University of Hong Kong, Pokfulam Road, Hong Kong}
\affiliation{The Hong Kong Institute for Astronomy and Astrophysics, The University of Hong Kong, Pokfulam Road, Hong Kong, P. R. China.}

\begin{abstract}

A robust stellar initial mass function (IMF) is crucial in any studies related to star formation. However, the direct measurement of the stellar IMF is confined to the local universe, limited by the resolving power of telescopes. Recently, a new method for accessing the stellar IMF beyond the local universe has been developed. The observed detection rate of transient lensed stars -- individual, massive, thus luminous stars in strongly lensed galaxies that are temporarily detectable upon stellar microlensing -- can serve as a probe to break the IMF-star formation history degeneracy in studies utilizing spectral energy distribution fitting, hence providing a window to look at the IMF at a subsample of gravitationally lensed galaxies. In this proceeding, I summarize the contributed talk given at IAUS402 entitled the same as this contribution and highlight some key results, which currently show no evidence for a top-heavy IMF in $z \approx 1$ galaxies.

\end{abstract}


\maketitle

\section{Introduction}

One of the apparent problems in studying massive stars is their formation efficiency -- describing the relative number of stars with different initial masses at any coeval star formation, the stellar initial mass function (IMF), often characterized as $dN/dm \propto (\frac{m}{M_{\odot}})^{-\alpha}$ \footnote{or $dN/d\, \textrm{log}\, m \propto (\frac{m}{M_{\odot}})^{\Gamma}$, such that $\Gamma = \alpha + 1$. Here I retain the $\alpha$ parameter to align with the main literature included.}. While local (from the Milky Way, to neighbouring galaxies such as SMC, LMC and M31) measurements carried out on resolved photometry deduce a perhaps surprisingly similar high-mass ($M \gtrsim 1.4\,M_{\odot}$) slope of the IMF of $\alpha \approx 2.3$ (e.g., \citealp{Salpeter_1955, Kroupa_2001}, see also \citealp{Massey_2025} for a recent review), the lack of resolving power to carry out photometry for individual star clusters away from our local universe makes it impossible to have a direct measurement of the stellar IMF. The IMF is thus often assumed to be universal across cosmic time in extragalactic astronomy. This assumption, however, might contribute to some of the mismatch between observations and predictions. One of the most famous examples is the number count of high-redshift massive galaxies observed by the {\it James-Webb Space Telescope} ({\it JWST}) \citep[e.g., ][]{Harikane_2023}. Recent works proposed that a top-heavy IMF, i.e., a shallower IMF that forms more massive stars than the local IMF, can resolve such a tension without appealing to the standard lambda CDM cosmology \citep[e.g., ][]{ziegler2025explainingtoomassivehighredshift}.

An emerging method in measuring the IMF beyond the local universe is by looking at the detection rate of transient lensed stars -- individual, luminous stars in gravitationally lensed background galaxies that are made temporarily detectable via stellar microlensing. First discovered less than a decade ago \citep{Kelly_2018_Icarus}, these transient lensed stars are now regularly detected in deep {\it Hubble Space Telescope} ({\it HST}) and {\it JWST} repeated observations of galaxy clusters \citep[e.g., ][]{Rodney_2018, Chen_2019_Warhol, Yan_2023, Meena_2023_Flashlights, Kelly_2022_Flashlights, Fudamoto_2025}. Given their cosmological distances ($z \gtrsim 1$), these stars must be intrinsically luminous (absolute magnitude, $M \lesssim -5$) to be detected even with the extreme magnification of up to $\sim 10^{4}$ \citep{Oguri_2018, Diego_2019_extrememagnification, Li_2025_CosmicHorseshoe}. These stars must therefore be (and only be) the most massive stars among the lensed stellar population, whose abundance depends on both (i) the star formation history (SFH), and (ii) the slope of the stellar IMF. While the IMF is well-known to be degenerate with the SFH \citep[e.g., ][]{Hoversten_2008}, the abundance of massive stars, thus the transient detection rate, is not. They hence become a unique probe to break the IMF-SFH degeneracy, and allow to distinguish the IMF where the SFH is constrained by spectral energy distribution (SED), and/or spectral features.

\section{Lensed stars' insight on the IMF}

Here, I discuss a few cases where insights on the high-mass end of the stellar IMF are placed by transient lensed star observations.

\subsection{``Spock''}

\begin{figure}[h!]
    \centering
    \includegraphics[width=\linewidth]{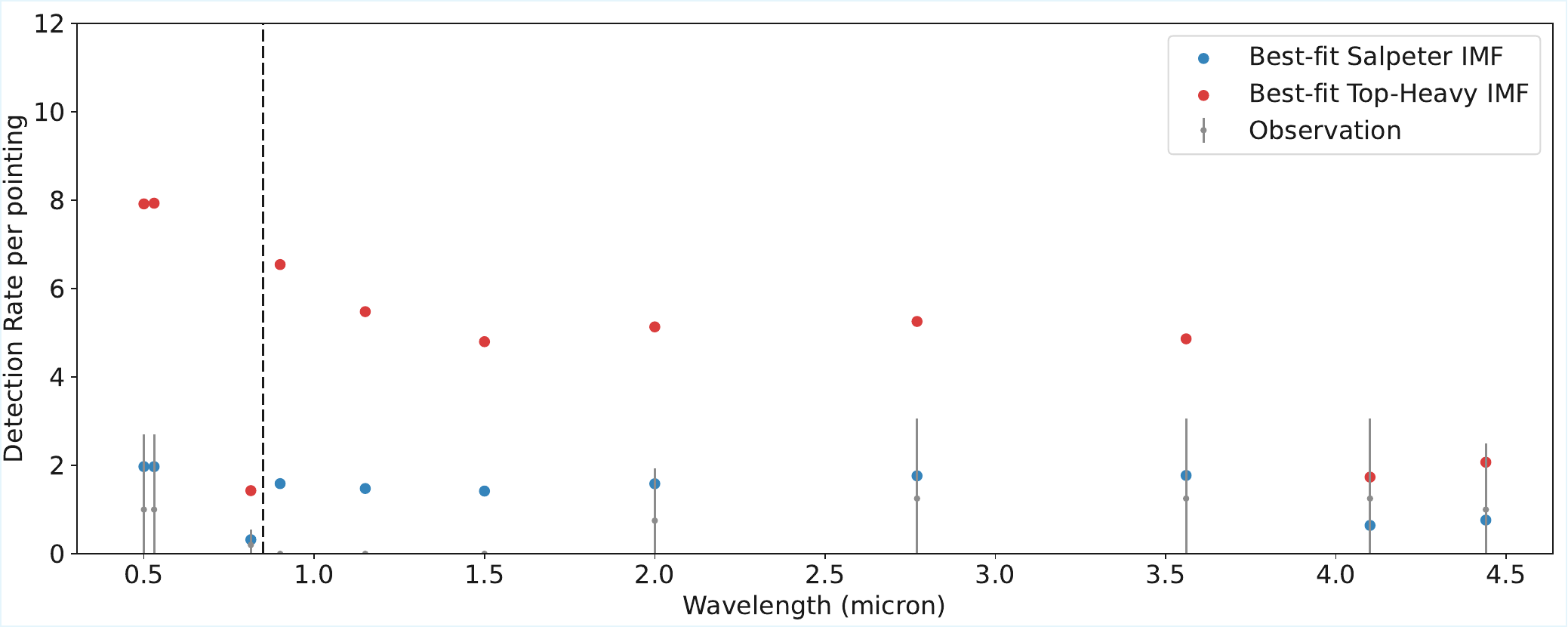}
    \caption{Observed transient detection rate in the Spock arc (gray data points), compared with predicted detection rate when adopting a Salpeter IMF (blue, $\alpha = 2.3$), and a top-heavy IMF ($\alpha = 1$, red) as predicted by \citet{Li_2025_IMF}. Filters leftward of the vertical dashed line are {\it HST} filters, whereas those on the right are {\it JWST} filters. }
    \label{fig: spock}
\end{figure}

``Spock'' ($z = 1.0$) is a galaxy strongly lensed by the foreground cluster MACSJ0416 ($z = 0.4$). Nine transient lensed star events have been discovered in this galaxy so far -- of which four of them are detected by {\it HST} \citep[thus are likely to be blue supergiants given their restframe UV brightness,][]{Rodney_2018, Kelly_2022_Flashlights}, and five of them are detected by {\it JWST} \citep[thus are likely to be red supergiants given their restframe optical/IR brightness,][]{Yan_2023, Williams_2025}.  
\citet{Li_2025_IMF} carried out an in-depth analysis of this arc, and compared the transient detection rate predicted using a Salpeter IMF (blue in Figure~\ref{fig: spock}), versus that using a Top-heavy IMF ($\alpha = 1.0$, red in Figure~\ref{fig: spock}) with SFH constrained by {\it HST} and {\it JWST} photometry. From Figure~\ref{fig: spock}, one can clearly see that the calculation adopting a shallower IMF significantly overpredicts the observed transient detection rate (gray) in almost all filters (except at $\gtrsim 4\,$ microns). Marginalizing over the choice of lens model and SFH models, the likelihood ratio analysis of \citet{Li_2025_IMF} indicates that given the model assumptions, the observation definitively favour the Spock galaxy to have a Salpeter-like IMF rather than a shallower top-heavy IMF.


\subsection{``Warhol''}

\begin{figure}
    \centering
    \includegraphics[width=0.7\linewidth]{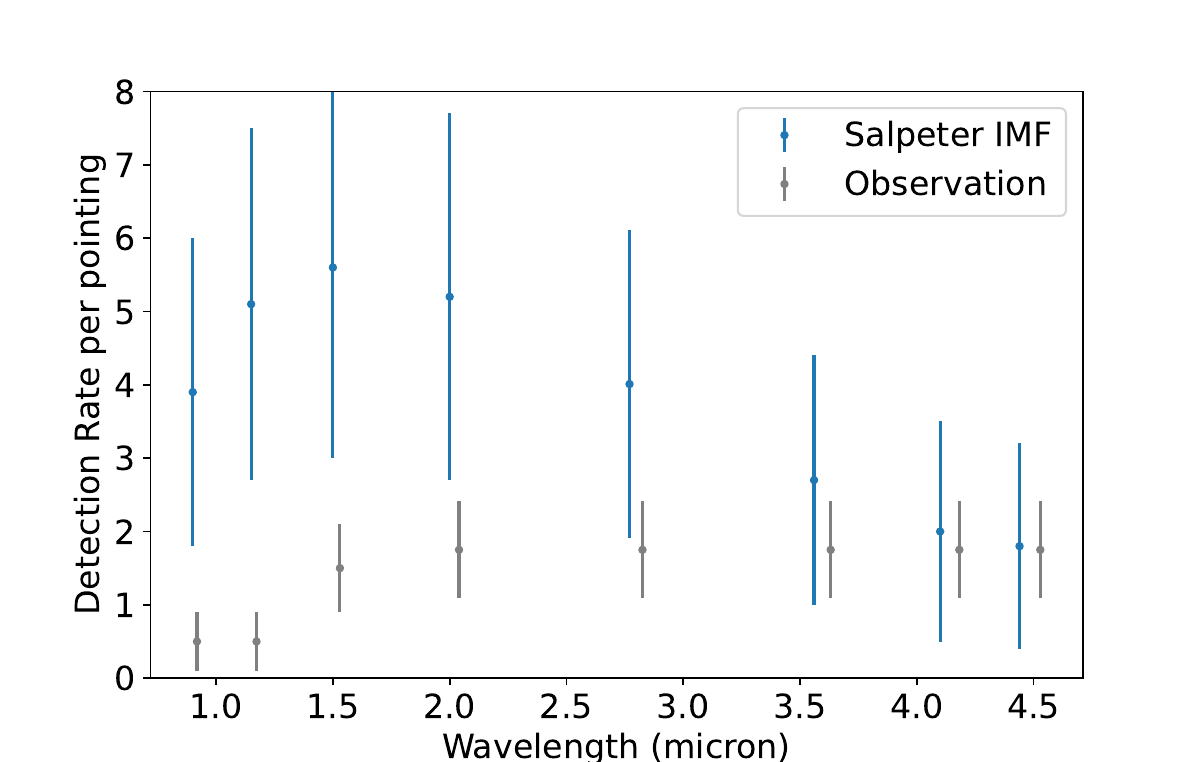}
    \caption{Observed transient detection rate (gray) versus transient detection rate predicted with a Salpeter IMF (blue) in the Warhol arc, numbers adopted from \citet{Palencia_2025_microlensing}.}
    \label{fig: warhol}
\end{figure}

``Warhol'' is another lensed galaxy in the same cluster, MACSJ0416, at a slightly lower redshift of $z = 0.94$ \citep{Chen_2019_Warhol}. \citet{Palencia_2025_microlensing} predicted the transient detection rate, similarly, for this galaxy. Although the initial purpose of this paper was not about the IMF, the fact that adopting a locally inferred IMF can reproduce the {\it JWST} detection rate \citep[nine transients over 4 visits,][]{Yan_2023, Williams_2025} down to an accuracy of $\sim 2\sigma$, as shown in Figure~\ref{fig: warhol}, indicates that there are no reasons to believe that this galaxy in particular, has an IMF deviates significantly from the locally measured value. Moreover, the detection rate when adopting a Salpeter-like IMF is already an overprediction -- meaning that a Top-heavy IMF is qualitatively not favoured by the data, as it would otherwise create an even stronger tension with observations. The consistency with a Salpeter IMF here further strengthens the case for such an IMF to be valid at $z \approx 1$.

\subsection{HST ``Flashlights'' Sample}

The two previous studies made use of data combining both {\it HST} and {\it JWST} observations, limited to two $z\approx1$ galaxies with many transient detections. There is a wider span of lensed arcs, at different redshifts, where their transient detection rate (albeit much lower) can also provide insights into the IMF. \citet{Meena_2025_IMF} carried a similar study (with a slightly different method in estimating the lensing effect), in which the detection rate of transients is marginalized over $\sim 10$ arcs (with a redshift range between $0.7 < z < 2.8$) across the six Hubble Frontier Field clusters, as observed by the {\it HST} Flashlights program \citep{Kelly_2022_Flashlights}. By marginalizing the detection rate over all these arcs, the observed detection rate of $20\pm5$, which falls between the predicted rate of simulations adopting $\alpha = 2.3$ ($\sim 13 \pm 4$) and $\alpha = 1$ ($\sim 35 \pm 6$) -- hinting that perhaps the IMF has evolved at a certain redshift. Combining this result with the conclusion in \citet{Li_2025_IMF} and \citet{Palencia_2025_microlensing}, the deviation of the IMF might have occurred at $z > 1$ during the cosmic noon.

\section{Conclusion}

Using lensing star transients to study the IMF is a very young field, yet it has demonstrated great potential in probing various astrophysical questions \citep[e.g., ][]{Kelly_2018_Icarus, Diego_2024_3M, Diego_2024_buffaloflashlights, Broadhurst_2025, Li_2025_BRratio}. Featuring individual, massive stars at cosmological distances, here in this proceeding, I highlight lensed stars' capability in probing the stellar IMF beyond the local universe, demonstrated in recent works \citep{Li_2025_IMF, Meena_2025_IMF, Palencia_2025_microlensing}. Future {\it JWST} observations have already been scheduled, and it is only a matter of time before we have a large number of detections of lensed stars. Finding no evidence of redshift $\sim 1$ galaxies having an IMF that deviates from the local measurement so far, extending the study to higher redshifts can allow one to trace further the validity of the IMF universality and allow for a better understanding of how the IMF might have evolved over cosmic time.  
\section*{Acknowledgment}

S.K.L. acknowledges the support from the Research Grants Council (RGC) of Hong Kong through the General Research Fund (GRF) 17302023.


\bibliography{Sample}{}

@misc{Massey_2025,
       author = {{Massey}, Philip and {Morrell}, Nidia I. and {Neugent}, Kathryn F. and {Herzog}, Monica and {Drout}, Maria R. and {O'Brien}, Caitlin},
        title = "{The Stellar Content of NGC\raisebox{-0.5ex}\textasciitilde3603 Revisited: Is the IMF Top Heavy?}",
      journal = {arXiv e-prints},
         year = 2025,
        month = sep,
          eid = {arXiv:2509.25099},
        pages = {arXiv:2509.25099},
          doi = {10.48550/arXiv.2509.25099},
archivePrefix = {arXiv},
       eprint = {2509.25099},
 primaryClass = {astro-ph.SR},
       adsurl = {https://ui.adsabs.harvard.edu/abs/2025arXiv250925099M},
      volume = {}
}

@misc{Li_2025_CosmicHorseshoe,
       author = {{Li}, Sung Kei and {Weisenbach}, Luke and {Collett}, Thomas E. and {Diego}, Jose M. and {Lim}, Jeremy and {Broadhurst}, Thomas J. and {Chow}, Alex and {Enzi}, Wolfgang J.~R. and {Kelly}, Patrick L. and {Melo-Carneiro}, Carlos R. and {Palencia}, Jose M. and {Williams}, Liliya L.~R. and {Zhang}, Jiashuo},
        title = "{Lensed stars in galaxy-galaxy strong lensing -- a JWST prediction for the Cosmic Horseshoe}",
      journal = {arXiv e-prints},
         year = 2025,
        month = sep,
          eid = {arXiv:2509.16154},
        pages = {arXiv:2509.16154},
          doi = {10.48550/arXiv.2509.16154},
archivePrefix = {arXiv},
       eprint = {2509.16154},
 primaryClass = {astro-ph.CO},
       adsurl = {https://ui.adsabs.harvard.edu/abs/2025arXiv250916154L},
      volume = {}
}

@misc{Li_2025_BRratio,
       author = {{Li}, Sung Kei and {Palencia}, Jose M. and {Diego}, Jose M. and {Lim}, Jeremy and {Kelly}, Patrick L. and {Meena}, Ashish K. and {Nianias}, James and {Williams}, Hayley and {Williams}, Liliya L.~R. and {Zitrin}, Adi},
        title = "{Transient star B/R ratio and star formation in $zrsim 1$ lensed galaxies}",
      journal = {arXiv e-prints},
         year = 2025,
        month = jun,
          eid = {arXiv:2506.17565},
        pages = {arXiv:2506.17565},
          doi = {10.48550/arXiv.2506.17565},
archivePrefix = {arXiv},
       eprint = {2506.17565},
 primaryClass = {astro-ph.CO},
       adsurl = {https://ui.adsabs.harvard.edu/abs/2025arXiv250617565L},
      volume = {}
}

@misc{Williams_2025,
       author = {{Williams}, Hayley and {Kelly}, Patrick L. and {Windhorst}, Rogier A. and {Filippenko}, Alexei V. and {Alfred}, Amruth and {Broadhurst}, Tom and {Chen}, Wenlei and {Conselice}, Christopher J. and {Cohen}, Seth H. and {Diego}, Jose M. and {Holwerda}, Benne W. and {Koekemoer}, Anton M. and {Li}, Sung Kei and {Meena}, Ashish Kumar and {Palencia}, Jose M. and {Ricotti}, Massimo and {Robertson}, Clayton D. and {Sun}, Bangzheng and {Willner}, S.~P. and {Yan}, Haojing and {Zitrin}, Adi},
        title = "{JWST's PEARLS: Temperatures of Nine Highly Magnified Stars in a Galaxy at Redshift z = 0.94 and Simulated Stellar Population Dependence on Stellar Metallicity and the Initial Mass Function}",
      journal = {arXiv e-prints},
         year = 2025,
        month = jul,
          eid = {arXiv:2507.03097},
        pages = {arXiv:2507.03097},
          doi = {10.48550/arXiv.2507.03097},
archivePrefix = {arXiv},
       eprint = {2507.03097},
 primaryClass = {astro-ph.GA},
       adsurl = {https://ui.adsabs.harvard.edu/abs/2025arXiv250703097W}
}

@misc{Broadhurst_2025,
       author = {{Broadhurst}, Tom and {Li}, Sung Kei and {Alfred}, Amruth and {Diego}, Jose M. and {Morilla}, Paloma and {Kelly}, Patrick L. and {Sun}, Fengwu and {Oguri}, Masamune and {Williams}, Hayley and {Windhorst}, Rogier and {Zitrin}, Adi and {Abe}, Katsuya T. and {Chen}, Wenlei and {Dai}, Liang and {Fudamoto}, Yoshinobu and {Kawai}, Hiroki and {Lim}, Jeremy and {Liu}, Tao and {Meena}, Ashish K. and {Palencia}, Jose M. and {Smoot}, George F. and {Williams}, Liliya L.~R.},
        title = "{Dark Matter Distinguished by Skewed Microlensing in the ``Dragon Arc''}",
      journal = {\apjl},
         year = 2025,
        month = jan,
       volume = {978},
       number = {1},
          eid = {L5},
        pages = {L5},
          doi = {10.3847/2041-8213/ad9aa8},
archivePrefix = {arXiv},
       eprint = {2405.19422},
 primaryClass = {astro-ph.CO},
       adsurl = {https://ui.adsabs.harvard.edu/abs/2025ApJ...978L...5B}
}

@misc{Fudamoto_2025,
       author = {{Fudamoto}, Yoshinobu and {Sun}, Fengwu and {Diego}, Jose M. and {Dai}, Liang and {Oguri}, Masamune and {Zitrin}, Adi and {Zackrisson}, Erik and {Jauzac}, Mathilde and {Lagattuta}, David J. and {Egami}, Eiichi and {Iani}, Edoardo and {Windhorst}, Rogier A. and {Abe}, Katsuya T. and {Bauer}, Franz Erik and {Bian}, Fuyan and {Bhatawdekar}, Rachana and {Broadhurst}, Thomas J. and {Cai}, Zheng and {Chen}, Chian-Chou and {Chen}, Wenlei and {Cohen}, Seth H. and {Conselice}, Christopher J. and {Espada}, Daniel and {Foo}, Nicholas and {Frye}, Brenda L. and {Fujimoto}, Seiji and {Furtak}, Lukas J. and {Golubchik}, Miriam and {Hsiao}, Tiger Yu-Yang and {Jolly}, Jean-Baptiste and {Kawai}, Hiroki and {Kelly}, Patrick L. and {Koekemoer}, Anton M. and {Kohno}, Kotaro and {Kokorev}, Vasily and {Li}, Mingyu and {Li}, Zihao and {Lin}, Xiaojing and {Magdis}, Georgios E. and {Meena}, Ashish K. and {Niemiec}, Anna and {Nabizadeh}, Armin and {Richard}, Johan and {Steinhardt}, Charles L. and {Wu}, Yunjing and {Zhu}, Yongda and {Zou}, Siwei},
        title = "{Identification of more than 40 gravitationally magnified stars in a galaxy at redshift 0.725}",
      journal = {Nature Astronomy},
         year = 2025,
        month = mar,
       volume = {9},
        pages = {428-437},
          doi = {10.1038/s41550-024-02432-3},
archivePrefix = {arXiv},
       eprint = {2404.08045},
 primaryClass = {astro-ph.GA},
       adsurl = {https://ui.adsabs.harvard.edu/abs/2025NatAs...9..428F}
}

@misc{Salpeter_1955,
       author = {{Salpeter}, Edwin E.},
        title = "{The Luminosity Function and Stellar Evolution.}",
      journal = {\apj},
         year = 1955,
        month = jan,
       volume = {121},
        pages = {161},
          doi = {10.1086/145971},
       adsurl = {https://ui.adsabs.harvard.edu/abs/1955ApJ...121..161S}
}

@misc{Li_2025_IMF,
       author = {{Li}, Sung Kei and {Diego}, Jose M. and {Meena}, Ashish K. and {Lim}, Jeremy and {Fung}, Leo W.~H. and {Levitskiy}, Arsen and {Nianias}, James and {Palencia}, Jose M. and {Williams}, Hayley and {Zhang}, Jiashuo and {Amruth}, Alfred and {Broadhurst}, Thomas J. and {Chen}, WenLei and {Filippenko}, Alexei V. and {Kelly}, Patrick L. and {Koekemoer}, Anton M. and {Perera}, Derek and {Sun}, Bangzheng and {Williams}, Liliya L.~R. and {Windhorst}, Rogier A. and {Yan}, Haojin and {Zitrin}, Adi},
        title = "{Constraining the z {\ensuremath{\sim}} 1 Initial Mass Function with HST and JWST Lensed Stars in MACS J0416.1‑2403}",
      journal = {\apj},
         year = 2025,
        month = aug,
       volume = {988},
       number = {2},
          eid = {178},
        pages = {178},
          doi = {10.3847/1538-4357/ade4bd},
       adsurl = {https://ui.adsabs.harvard.edu/abs/2025ApJ...988..178L}
}

@misc{Palencia_2025_microlensing,
       author = {{Palencia}, J.~M. and {Diego}, J.~M. and {Dai}, L. and {Pascale}, M. and {Windhorst}, R. and {Koekemoer}, A.~M. and {Li}, Sung Kei and {Kavanagh}, B.~J. and {Sun}, Fengwu and {Alfred}, Amruth and {Meena}, Ashish K. and {Broadhurst}, Thomas J. and {Kelly}, Patrick L. and {Perera}, Derek and {Williams}, Hayley and {Zitrin}, Adi},
        title = "{Microlensing at cosmological distances: Event rate predictions in the Warhol arc of MACS 0416}",
      journal = {AAP},
         year = 2025,
        month = jul,
       volume = {699},
          eid = {A295},
        pages = {A295},
          doi = {10.1051/0004-6361/202555447},
archivePrefix = {arXiv},
       eprint = {2504.07039},
 primaryClass = {astro-ph.CO},
       adsurl = {https://ui.adsabs.harvard.edu/abs/2025A&A...699A.295P}
}

@misc{Diego_2024_buffaloflashlights,
       author = {{Diego}, Jose M. and {Li}, Sung Kei and {Meena}, Ashish K. and {Niemiec}, Anna and {Acebron}, Ana and {Jauzac}, Mathilde and {Struble}, Mitchell F. and {Amruth}, Alfred and {Broadhurst}, Tom J. and {Cerny}, Catherine and {Ebeling}, Harald and {Filippenko}, Alexei V. and {Jullo}, Eric and {Kelly}, Patrick and {Koekemoer}, Anton M. and {Lagattuta}, David and {Lim}, Jeremy and {Limousin}, Marceau and {Mahler}, Guillaume and {Patel}, Nency and {Remolina}, Juan and {Richard}, Johan and {Sharon}, Keren and {Steinhardt}, Charles and {Umetsu}, Keiichi and {Williams}, Liliya and {Zitrin}, Adi and {Palencia}, Jose Mar{\'\i}a and {Dai}, Liang and {Ji}, Lingyuan and {Pascale}, Massimo},
        title = "{BUFFALO/Flashlights: Constraints on the abundance of lensed supergiant stars in the Spock galaxy at redshift 1}",
      journal = {aap},
         year = 2024,
        month = jan,
       volume = {681},
          eid = {A124},
        pages = {A124},
          doi = {10.1051/0004-6361/202346761},
archivePrefix = {arXiv},
       eprint = {2304.09222},
 primaryClass = {astro-ph.GA},
       adsurl = {https://ui.adsabs.harvard.edu/abs/2024A&A...681A.124D}
}

@misc{Meena_2025_IMF,
       author = {{Meena}, Ashish Kumar and {Li}, Sung Kei and {Zitrin}, Adi and {Kelly}, Patrick L. and {Broadhurst}, Tom and {Chen}, Wenlei and {Diego}, Jose M. and {Filippenko}, Alexei V. and {Furtak}, Lukas J. and {Williams}, Liliya L.~R.},
        title = "{Flashlights: Prospects for constraining the initial mass function around cosmic noon with caustic-crossing events}",
      journal = {aap},
         year = 2025,
        month = jul,
       volume = {699},
          eid = {A299},
        pages = {A299},
          doi = {10.1051/0004-6361/202555023},
archivePrefix = {arXiv},
       eprint = {2503.21706},
 primaryClass = {astro-ph.GA},
       adsurl = {https://ui.adsabs.harvard.edu/abs/2025A&A...699A.299M}
}

@misc{Diego_2024_3M,
       author = {{Diego}, Jose M. and {Li}, Sung Kei and {Amruth}, Alfred and {Meena}, Ashish K. and {Broadhurst}, Tom J. and {Kelly}, Patrick L. and {Filippenko}, Alexei V. and {Williams}, Liliya L.~R. and {Zitrin}, Adi and {Harris}, William E. and {Reina-Campos}, Marta and {Giocoli}, Carlo and {Dai}, Liang and {Struble}, Mitchell F. and {Treu}, Tommaso and {Fudamoto}, Yoshinobu and {Gilman}, Daniel and {Koekemoer}, Anton M. and {Lim}, Jeremy and {Palencia}, Jose Mar{\'\i}a and {Sun}, Fengwu and {Windhorst}, Rogier A.},
        title = "{Imaging dark matter at the smallest scales with z {\ensuremath{\approx}} 1 lensed stars}",
      journal = {\aap},
         year = 2024,
        month = sep,
       volume = {689},
          eid = {A167},
        pages = {A167},
          doi = {10.1051/0004-6361/202450474},
archivePrefix = {arXiv},
       eprint = {2404.08033},
 primaryClass = {astro-ph.CO},
       adsurl = {https://ui.adsabs.harvard.edu/abs/2024A&A...689A.167D}
}
\bibliographystyle{iaulike}


\end{document}